# SIR models with demography, random transmission coefficient and non-autonomous vaccination rate

Javier López-de-la-Cruz[a,1], Susana Merchán[b], Felipe Rivero[c] and Javier Rodrigo[d]

[a] Dpto. de Matemática Aplicada a las TIC, Escuela Técnica Superior de Ingenieros Informáticos, Campus de Montegancedo, Universidad Politécnica de Madrid, 28660 Boadilla del Monte, Madrid, Spain.
[b] Dpto. de Matemática e Informática Aplicadas a las Ingenierías Civil y Naval, Escuela Técnica Superior de Ingenieros de Caminos, Canales y Puertos, Campus de Ciudad Universitaria, Universidad Politécnica de Madrid, 28040, Madrid, Spain.
[c] Dpto. de Matemática Aplicada a las TIC, Escuela Técnica Superior de Ingeniería y Sistemas de Telecomunicación, Campus Sur, Universidad Politécnica de Madrid, 28031, Madrid, Spain.
[d] Departamento de Matemática Aplicada, Escuela Técnica Superior de Ingeniería, Universidad Pontificia Comillas de Madrid, 28015, Madrid, Spain

**Abstract**

In this paper we investigate the asymptotic behavior of some SIR models incorporating demography, bounded random transmission coefficient and a time-dependent vaccination strategy targeting the susceptible population. In this setting, we establish the existence and uniqueness of non-negative global solution of the models and derive conditions under which either the disease is eradicated or becomes endemic. In addition, the theoretical results are further illustrated by several numerical simulations.

## 1. Introduction

Throughout history, epidemics have profoundly influenced human societies, often producing consequences as severe as those arising from armed conflicts and, in some cases, leading to the extinction of entire populations. Beyond the tragic human cost, epidemics also generate significant social disruption and exert a considerable negative impact on economic systems (see [11, 18]).

Many infectious diseases that have historically triggered epidemics remain uneradicated, and new pathogens continue to emerge over time. This reality underscores the critical need for the development of mathematical models that can accurately characterize the transmission of infectious diseases. Such models not only enhance our understanding

[1] corresponding author: javier.lopez.delacruz@upm.es

of disease propagation but also support timely and effective decision-making aimed at mitigating the spread and reducing the adverse consequences of future outbreaks.

The earliest known contribution to the mathematical modeling of epidemics dates back to 1760, when D. Bernoulli employed a system of ordinary differential equations to analyze the dissemination of smallpox (see [3]). Nevertheless, it was not until the early 20th century that the formal study of mathematical models in epidemiology began to advance.

In 1906, W. H. Hamer introduced a discrete-time model to describe measles transmission (see [12]). Subsequently, in 1911, R. Ross formulated a mathematical model to predict the spread of malaria (see [19]), demonstrating that a partial reduction in the mosquito population could suffice to eradicate the disease.

A particularly significant milestone was achieved in 1927, when Kermack and McKendrick introduced a mathematical framework for modeling the spread of infectious diseases (see [15]). Their model, widely known as the SIR model, has since become the foundational structure for mathematical epidemiology.

This compartmental model categorizes the population into three distinct groups: the susceptible ($S$), consisting of individuals who are at risk of infection upon contact with an infectious person; the infected ($I$), representing individuals actively carrying and transmitting the pathogen; and the recovered ($R$), comprising individuals who have either developed immunity following recovery or succumbed to the disease.

Figure 1 illustrates the state-transition diagram associated with the classical SIR model, where $\beta > 0$ (in days$^{-1}$) represents the transmission coefficient, quantifying the rate at which the disease spreads through contact between susceptible and infected individuals, and $1/\delta > 0$ (in days) defines the average infectious period, that is, the expected time an individual remains capable of transmitting the disease following infection.

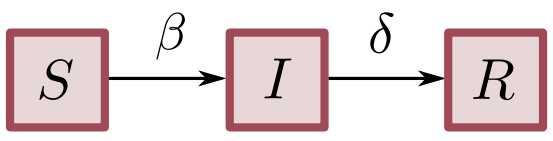

Figure 1: State-transition diagram of the classical SIR model.

As depicted in the diagram, individuals in the susceptible compartment transition to the infected state exclusively upon acquiring the infection, typically as a result of direct interaction with an infected individual. Additionally, individuals in the infected compartment exit this state either upon recovery, thereby acquiring immunity, or as a consequence of disease-induced mortality.

The classical SIR model proposed by Kermack and McKendrick is given by the following nonlinear differential system

$$S'(t) = -\beta \frac{S(t)I(t)}{S(t)+I(t)+R(t)}, \tag{1}$$

$$I'(t) = \beta \frac{S(t)I(t)}{S(t)+I(t)+R(t)} - \delta I(t), \tag{2}$$

$$R'(t) = \delta I(t), \tag{3}$$

where $S = S(t)$, $I = I(t)$ and $R = R(t)$ denote the number of susceptible, infected and recovered individuals, respectively, at time $t$ (in days).

Even though the classical SIR model (1)-(3) has been widely studied in the literature by scientists from many areas of knowledge, it is not enough to represent in a loyal manner the evolution of real pandemics as the COVID-19, since it does not take into consideration some essential ingredients that can be easily observed in real life.

On the one hand, it is important to remark that the classical SIR model (1)-(3) can be used to study epidemics in shorts periods of time, but it is not suitable to describe the evolution of epidemics that last a long time as in the case of the COVID-19, since some new people typically join the population or might leave the population (due to natural deaths or others, for instance) during that period of time.

To solve this drawback, we introduce below the classical SIR model with demography

$$S'(t) = \nu - \beta \frac{S(t)I(t)}{S(t)+I(t)+R(t)} - \mu S(t), \tag{4}$$

$$I'(t) = \beta \frac{S(t)I(t)}{S(t)+I(t)+R(t)} - \delta I(t) - \mu I(t), \tag{5}$$

$$R'(t) = \delta I(t) - \mu R(t), \tag{6}$$

where $\nu$ (in number of individuals $\times$ days$^{-1}$) represents the recruitment rate of the population and $\mu$ (in days$^{-1}$) corresponds to the mortality rate of the population.

The state-transition diagram associated to the classical SIR model with demography (4)-(6) is given in Figure 2.

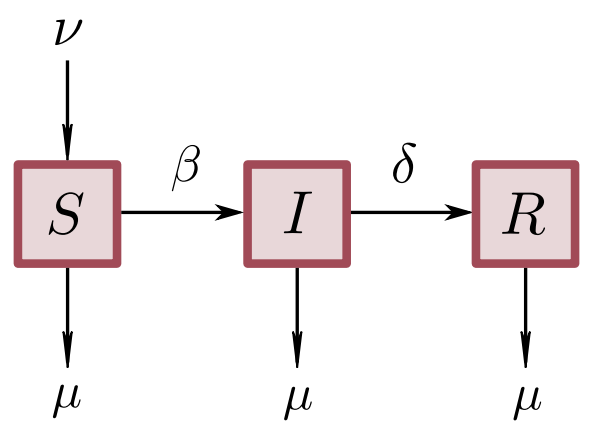

Figure 2: State-transition diagram of the classical SIR model with demography.

On the other hand, neither the classical SIR model (1)-(3) nor the classical SIR model with demography (4)-(6) take into consideration the possible vaccination of individuals. It is well known that vaccines have radically changed the course of COVID-19 and, then, this ingredient should not be be omitted when developing realistic models.

Then, we propose in this paper a new SIR model, in which the vaccination of susceptible individuals is taken into account (this choice is based on the fact that this is the most common vaccination strategy and the one carried out to combat the COVID-19), given as

$$S'(t) = \nu - \beta \frac{S(t)I(t)}{S(t)+I(t)+R(t)} - \mu S(t) - pS(t), \tag{7}$$

$$I'(t) = \beta \frac{S(t)I(t)}{S(t)+I(t)+R(t)} - \delta I(t) - \mu I(t), \tag{8}$$

$$R'(t) = \delta I(t) - \mu R(t) + pS(t), \tag{9}$$

where $p$ (in days$^{-1}$) denotes the proportion of susceptible individuals that are vaccinated.

Figure 3 shows the state-transition diagram associated to the classical SIR model with demography and vaccination of susceptible individuals (7)-(9).

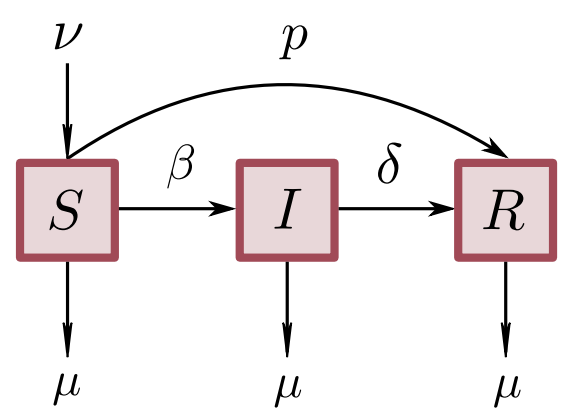

Figure 3: State-transition diagram of the classical SIR model with demography and vaccination of susceptible individuals.

Nevertheless, in this paper we go further and propose to introduce two new ingredients that are essential when modeling epidemics.

The first one consists of introducing randomness into the transmission coefficient of the disease. The SIR models presented up to now assume that the transmission coefficient of the disease is constant. However, this coefficient is very related to movements of people and movements of people are, obviously, subject to randomness (see [1]).

The second ingredient is based on considering a non-autonomous vaccination rate. Every SIR model introduced above supposes a constant vaccination rate. Nevertheless, it is much more realistic to consider the vaccination rate as time-dependent function. It could be random, but it is natural to assume it being non-autonomous since we can control it.

As a result, in this paper we propose to study the following SIR model with demography,

random transmission coefficient and non-autonomous vaccination rate

$$S'(t) = \nu - (\beta + \Phi(z^*(\theta_t\omega)) \frac{S(t)I(t)}{S(t)+I(t)+R(t)} - \mu S(t) - p(t)S(t), \tag{10}$$

$$I'(t) = (\beta + \Phi(z^*(\theta_t\omega)) \frac{S(t)I(t)}{S(t)+I(t)+R(t)} - \delta I(t) - \mu I(t), \tag{11}$$

$$R'(t) = \delta I(t) - \mu R(t) + p(t)S(t), \tag{12}$$

where $p : \mathbb{R} \to [0, \overline{p}]$, with $0 \leq \overline{p} \leq 1$, is the function describing how vaccines are administered and $\Phi(z^*(\theta_t\omega))$ denotes a bounded noise (see Section 2 for details).

The rest of the paper is organized as follows: in Section 2 we include some preliminaries about the Ornstein-Uhlenbeck process and the bounded noise. In Section 3 we prove the well-posedness of the SIR model with demography, random transmission coefficient and non-autonomous vaccination rate (10)-(12). After that, in Section 4 and Section 5 we analyze in detail the long-time dynamics of systems (7)-(9) and (10)-(12), respectively. We provide conditions under which the disease either is eradicated or becomes endemic and we depict numerical simulations to illustrate and support the theoretical results. Finally, in Section 6 we include some conclusions and final comments.

## 2. Preliminaries on the bounded noise

In this section we include some preliminaries about the Ornstein-Uhlenbeck process and the bounded noise $\Phi(z^*(\theta_t\omega))$, appearing in system (10)-(12), necessary to facilitate the understanding of the rest of the paper and make it as much self-contained as possible.

Let us start considering the probability space $(\Omega, \mathcal{F}, \mathbb{P})$, where $\Omega = \mathcal{C}^0(\mathbb{R}; \mathbb{R})$ denotes the space of continuous functions from $\mathbb{R}$ to itself being zero at zero, $\mathcal{F}$ is the Borel $\sigma-$algebra on $\Omega$ generated by the compact open topology (see Appendix A.2 and Appendix A.3 in [2]) and $\mathbb{P}$ the corresponding Wiener measure.

In addition, let $\{\theta_t\}_{t\in\mathbb{R}}$ be a family of mappings $\theta_t : \Omega \to \Omega$ defined as

$$\theta_t\omega(\cdot) = \omega(\cdot + t) - \omega(t), \quad t \in \mathbb{R},$$

which is known as Wiener shift flow and satisfies the following properties:

(1) $\theta_0 = \mathrm{Id}_\Omega$,

(2) $\theta_s \circ \theta_t = \theta_{s+t}$ for all $s, t \in \mathbb{R}$,

(3) the mapping $(t, \omega) \to \theta_t\omega$ is measurable, and

(4) the probability measure $\mathbb{P}$ is preserved by $\theta_t$ (i.e., $\theta_t\mathbb{P} = \mathbb{P}$) for all $t \in \mathbb{R}$.

Then, on $(\Omega, \mathcal{F}, \mathbb{P}, \{\theta_t\}_{t\in\mathbb{R}})$ we can define the Ornstein-Uhlenbeck process as follows

$$z(t,\omega) := z^*(\theta_t\omega) = -\int_{-\infty}^{0} e^s \theta_t\omega(s)ds, \tag{13}$$

for any $t \in \mathbb{R}$ and $\omega \in \Omega$, which is a mean-reverting stochastic process, i.e., the probability of the process to go back to its mean value increases when the process is far away from its mean value (see [9] for more details).

Then, we can consider now the mapping $\Phi : \mathbb{R} \to [-d, d]$, given by

$$\Phi(z) = \frac{2d}{\pi} \arctan(z), \tag{14}$$

where $d > 0$ is a constant typically provided by practitioners, and define a new stochastic process as $\Phi(z^*(\theta_t\omega)$ satisfying the properties summarized in the following proposition.

**Proposition 2.1.** *Let $\Phi$ be a function given as in* (14) *and consider $z^*(\theta_t\omega)$ the Ornstein-Uhlenbeck process. Then:*

*(1) the mapping $t \to \Phi(z^*(\theta_t\omega)$ is continuous for almost every $\omega \in \Omega$, and*

*(2) the following property*

$$\lim_{t\to+\infty} \frac{1}{t} \int_0^t \Phi(z^*(\theta_s\omega))ds = 0 \tag{15}$$

*fulfills for almost every $\omega \in \Omega$.*

The proof of the first statement of Proposition 2.1 can be found in [2, 9]. Concerning the second statement, it is proved in [5, Theorem 4.1].

**Remark 2.1.** *The second statement of Proposition 2.1 holds true for every function $\Phi$ being continuous, odd and bounded.*

**Remark 2.2.** *The reader is referred to [6, 7, 10, 16, 20] for an in-depth treatment of the modeling of bounded stochastic perturbations via the Ornstein-Uhlenbeck process, not only in the context of mathematical epidemiology but also within a broader range of population dynamics frameworks.*

## 3. Well-posedness of the SIR model with demography, random transmission coefficient and non-autonomous vaccination rate

In this section, we prove that the SIR model with demography, random transmission coefficient and non-autonomous vaccination rate, given by system (10)-(12), is well-posed, that is, it admits a unique solution, which is defined globally in time and remains non-negative for any non-negative initial condition.

Let us define $\mathbb{R}^3_+ = \{(S, I, R) \in \mathbb{R}^3 : S \geq 0, I \geq 0, R \geq 0\}$, the first octant in $\mathbb{R}^3$.

**Theorem 3.1.** *For every initial condition* $u_0 = (S_0, I_0, R_0) \in \mathbb{R}^3_+$ *and* $\omega \in \Omega$*, system* (10)-(12) *admits a unique solution*

$$u(t; 0, \omega, u_0) = ((S(t; 0, \omega, u_0), I(t; 0, \omega, u_0), R(t; 0, \omega, u_0)) \in \mathcal{C}^1([0, +\infty); \mathbb{R}^3_+),$$

*where* $u(t; 0, \omega, u_0)$ *denotes the value at time* $t$ *of the solution of system* (10)-(12) *depending on* $\omega$ *and starting with initial condition* $u_0$ *at time* 0*. In addition,* $S_0 = S(0; 0, \omega, u_0)$*,* $I_0 = I(0; 0, \omega, u_0)$ *and* $R_0 = R(0; 0, \omega, u_0)$*.*

*Proof.* Since, the Ornstein-Uhlenbeck process has continuous trajectories and $\Phi$ is continuous, the right-hand side of system (10)-(12) is continuous with respect to $t$. Moreover, the vector field of system (10)-(12) is $\mathcal{C}^1(\mathbb{R}^3)$ with respect to $(S, I, R)$, whence it is locally Lipschitz with respect to $(S, I, R)$. Thus, system (10)-(12) admits a unique local solution for every initial condition in $\mathbb{R}^3$, thanks to the classical theory of ordinary differential equations (see, for instance, [8, 21]).

Suppose now that there is a time $t^* \geq 0$ such that $S(t^*) = 0$, $I(t^*) \geq 0$ and $R(t^*) \geq 0$. Hence, from (10), $S'(t^*) = \nu > 0$, then $S$ is increasing at $t^*$ and $S$ can not take negative values. Let assume now that $I(t^*) = 0$, $S(t^*) \geq 0$ and $R(t^*) \geq 0$ for some time $t^* \geq 0$. In this case, since $I = 0$ solves (11), it follows trivially that $I$ can not take negative values due to the uniqueness of local solution. Finally, if $R(t^*) = 0$, $S(t^*) \geq 0$ and $I(t^*) \geq 0$ for some $t^* \geq 0$, it follows from (12) that $R'(t^*) = \delta I(t^*) + p(t^*)S(t^*) \geq 0$, whence $R$ is non-decreasing at $t^*$ and then $R$ can not take negative values. This proves that every solution of system (10)-(12) with initial condition in $\mathbb{R}^3_+$ remains in $\mathbb{R}^3_+$.

Define now $N(t) = S(t) + I(t) + R(t)$, the total population at time $t$, which satisfies

$$N'(t) = \nu - \mu N(t), \tag{16}$$

whose solution, for every initial data $N_0 = S_0 + I_0 + R_0$ and $\omega \in \Omega$, is given by

$$N(t; 0, \omega, N_0) = N_0 e^{-\mu t} + \frac{\nu}{\mu}\left(1 - e^{-\mu t}\right) \tag{17}$$

for every $t \geq 0$.

From (17), it is easy to notice that $N$ can not blow up at any finite time. This, jointly with the fact that both $S$, $I$ and $R$ are non-negative for every non-negative initial condition (and, then, $N$ is non-negative), allows us to deduce that $S$, $I$ and $R$ can not blow up at any finite time either, whence the unique local solution of system (10)-(12) is, in fact, a global one, i.e., it is defined for every $t \geq 0$. □

## 4. SIR model with demography and vaccination of susceptible individuals

In this section we investigate the autonomous SIR model with vital dynamics and vaccination of susceptible individuals given by system (7)-(9).

The existence and uniqueness of non-negative global solution of system (7)-(9) follows directly from Proposition 3.1, since system (7)-(9) is a particular case of system (10)-(12) with no noise, i.e., when $\Phi(z^*(\theta_t\omega)) \equiv 0$.

In the sequel we mainly focus on analyzing in detail the long-time behavior of system (7)-(9). More precisely, our main goal is to provide conditions on the parameters of system (7)-(9) under which the disease is eradicated or becomes endemic.

### 4.1. Long-time dynamics of the solutions

In what follows, we provide the equilibrium points of system (7)-(9) and study their local stability (see [8, 13] for more information about the theoretical framework).

**Proposition 4.1.** *System* (7)-(9) *possesses the equilibrium point*

$$p_1 = \left( \frac{\nu}{\mu+p}, 0, \frac{\nu p}{\mu(\mu+p)} \right), \tag{18}$$

*and a new one, given by*

$$p_2 = \left( \frac{\nu(\delta+\mu)}{\beta\mu}, \frac{\nu\left(\mu(\beta-\delta-\mu) - p(\delta+\mu)\right)}{\beta\mu(\delta+\mu)}, \frac{\nu(p(\delta+\mu) - \delta(-\beta+\delta+\mu))}{\beta\mu(\delta+\mu)} \right), \tag{19}$$

*arises in* $\mathbb{R}^3_+$ *as long as condition*

$$\beta > \frac{(\delta+\mu)(\mu+p)}{\mu} \tag{20}$$

*is satisfied.*

*Proof.* It is easy to check that both $p_1$ and $p_2$ are the solutions of the algebraic system

$$0 = \nu - \beta\frac{SI}{S+I+R} - \mu S - pS, \tag{21}$$

$$0 = -\beta\frac{SI}{S+I+R} - \delta I - \mu I, \tag{22}$$

$$0 = \delta I - \mu R + pS. \tag{23}$$

Moreover, $p_1 \in \mathbb{R}^3_+$ or, in other words, it has biological sense.

In addition, notice that

$$\frac{(\delta+\mu)(\mu+p)}{\mu} - \frac{(\delta+\mu)(\delta-p)}{\delta} = \frac{p(\delta+\mu)^2}{\delta\mu} > 0$$

whence

$$\frac{(\delta+\mu)(\mu+p)}{\mu} > \frac{(\delta+\mu)(\delta-p)}{\delta}.$$

Then, thanks to (20), we obtain that

$$\beta > \frac{(\delta+\mu)(\delta-p)}{\delta},$$

and this guarantees that $p_2 \in \mathbb{R}^3_+$, i.e., it also has biological sense. □

**Remark 4.1.** *Notice that* $p_1$ *corresponds to a disease-free equilibrium, where the infection is eradicated. Moreover, every component of* $p_2$ *is strictly positive if* (20) *fulfills, which represents a biological state in which the disease becomes endemic.*

In the following propositions we are going to study the local stability of both equilibrium points $p_1$ and $p_2$.

**Proposition 4.2.** *The equilibrium point* $p_1$ *is locally unstable as long as condition* (20) *is fulfilled. However,* $p_1$ *is locally asymptotically stable provided that condition*

$$\beta < \frac{(\delta+\mu)(\mu+p)}{\mu}, \tag{24}$$

*holds true.*

*Proof.* The Jacobian matrix of the right-hand of system (21)-(23) evaluated at $p_1$ is

$$\begin{pmatrix} -\mu-p & -\dfrac{\beta\mu}{\mu+p} & 0 \\ 0 & -\dfrac{\mu(-\beta+\delta+\mu)+p(\delta+\mu)}{\mu+p} & 0 \\ p & \delta & -\mu \end{pmatrix}, \tag{25}$$

whose eigenvalues are given by

$$\lambda_1 = -\mu-p, \quad \lambda_2 = -\frac{\mu(-\beta+\delta+\mu)+p(\delta+\mu)}{\mu+p}, \quad \text{and} \quad \lambda_3 = -\mu.$$

Both eigenvalues $\lambda_1$ and $\lambda_3$ are negative. Moreover, the numerator of $\lambda_2$ is negative if (20) is satisfied, whence the eigenvalue is positive. However, $\lambda_2$ is negative as long as condition (24) holds true. This proves the proposition. □

**Proposition 4.3.** *The equilibrium point* $p_2$ *is locally asymptotically stable if* (20) *fulfills.*

*Proof.* The Jacobian matrix of the right-hand of system (21)-(23) evaluated at $p_2$ is

$$\begin{pmatrix} -\dfrac{\beta\mu}{\delta+\mu}+\mu-\dfrac{(\delta+\mu)(\mu+p)}{\beta} & -\dfrac{\beta\delta+\mu(\delta+\mu)+p(\delta+\mu)}{\beta} & -\dfrac{\mu(-\beta+\delta+\mu)+p(\delta+\mu)}{\beta} \\ \dfrac{(-\beta+\delta+\mu)(\mu(-\beta+\delta+\mu)+p(\delta+\mu))}{\beta(\delta+\mu)} & \dfrac{\mu(-\beta+\delta+\mu)+p(\delta+\mu)}{\beta} & \dfrac{\mu(-\beta+\delta+\mu)+p(\delta+\mu)}{\beta} \\ p & \delta & -\mu \end{pmatrix},$$

whose eigenvalues are given as follows

$$\lambda_1 = -\mu, \quad \lambda_2 = -\frac{\beta\mu + \eta}{2(\delta + \mu)}, \quad \text{and} \quad \lambda_3 = -\frac{\beta\mu - \eta}{2(\delta + \mu)},$$

where

$$\eta = \sqrt{\mu\left(\beta^2\mu - 4\beta(\delta + \mu)^2 + 4(\delta + \mu)^3\right) + 4p(\delta + \mu)^3}.$$

The eigenvalue $\lambda_1$ is negative and the real part of the eigenvalue $\lambda_2$ is also negative. In addition, the real part of the eigenvalue $\lambda_3$ is negative whether (20) is satisfied. □

**Remark 4.2.** *Thanks to Proposition 4.2 and Proposition 4.3, we can define now the basic reproduction number of system* (7)-(9) *as*

$$\mathcal{R}_0 = \frac{\beta\mu}{(\delta + \mu)(\mu + p)}, \tag{26}$$

*which quantifies the average number of susceptible individuals that a single infected individual can transmit the disease during the infectious period. This number allows us to fully characterize, for $\mathcal{R}_0 \neq 1$, the dynamics of system* (7)-(9)*. Specifically, the disease will be eradicated whenever $\mathcal{R}_0 < 1$, while it will become endemic if $\mathcal{R}_0 > 1$.*

In the sequel, we analyze the long-time dynamics of the solutions of system (7)-(9).

**Theorem 4.1.** *Consider the compact set*

$$\mathcal{A} = \left\{ (S, I, R) \in \mathbb{R}^3_+ \,:\, S + I + R = \frac{\nu}{\mu} \right\}. \tag{27}$$

*(1) The set $\mathcal{A}$ is attracting for the solutions of system* (7)-(9)*, in the sense that*

$$\lim_{t\to+\infty} \sup_{u_0 \in F} \inf_{a \in \mathcal{A}} \|u(t; 0, u_0) - a\| = 0,$$

*where $F \subset \mathbb{R}^3_+$ denotes a bounded set where the initial conditions are considered.*

*(2) The set $\mathcal{A}$ is invariant for system* (7)-(9)*, i.e., every solution of system* (7)-(9) *with initial condition in $\mathcal{A}$ always remains in $\mathcal{A}$.*

*Proof.* Recall that the total population $N(t) = S(t) + I(t) + R(t)$ satisfies (16), whose solution verifies

$$\lim_{t\to+\infty} N(t; 0, N_0) = \frac{\nu}{\mu},$$

whence we deduce that $\mathcal{A}$, given by (27), is attracting for the solutions of system (7)-(9).

The invariance of $\mathcal{A}$ follows trivially since $N = \nu/\mu$ solves (16). □

In view of Theorem 4.1, the rest of this section is dedicated to investigate in detail the asymptotic dynamics of system (7)-(9) restricted to the attracting set $\mathcal{A}$ in (27).

**Theorem 4.2.** *The equilibrium point $p_2$ is globally asymptotically stable in $\mathcal{A}$ if* (20) *is satisfied. Otherwise, the equilibrium point $p_1$ is globally asymptotically stable in $\mathcal{A}$.*

*Proof.* We proceed by studying the nullclines (see [13, Chapter 9] for the theoretical framework) of system (7)-(9) restricted to $\mathcal{A}$, which is given by

$$S'(t) = \nu - \frac{\beta\mu}{\nu}S(t)I(t) - (\mu + p)S(t), \tag{28}$$

$$I'(t) = \left(\frac{\beta\mu}{\nu}S(t) - (\delta + \mu)\right) I(t). \tag{29}$$

It is not difficult to check the $I-$nullclines of system (28)-(29) are

$$I = 0, \tag{30}$$

$$S = \frac{\nu(\delta + \mu)}{\beta\mu}, \tag{31}$$

whereas system (28)-(29) possesses only the $S-$nullcline

$$I = \frac{\nu}{\beta\mu}\left(\frac{\nu}{S} - (\mu + p)\right). \tag{32}$$

Recall that $I' = 0$ on the $I-$nullclines (30)-(31) and $S' = 0$ on the $S-$nullcline (32). Thus, the nullclines divide $\mathcal{A}$ into open regions where $I'$ and $S'$ are either strictly positive or strictly negative, that is, where $I$ and $S$ are strictly increasing or decreasing.

In addition, the intersection of the $I-$nullcline (30) with the $S-$nullcline (32) gives

$$e_1 = \left(\frac{\nu}{\mu + p}, 0\right),$$

which is an equilibrium point of system (28)-(29) and corresponds with the equilibrium point $p_1$ of system (7)-(9) restricted to $\mathcal{A}$. Similarly, the intersection of the $I-$nullcline (31) with the $S-$nullcline (32) gives

$$e_2 = \left(\frac{\nu(\delta + \mu)}{\beta\mu}, \frac{\nu\left(\mu(\beta - \delta - \mu) - p(\delta + \mu)\right)}{\beta\mu(\delta + \mu)}\right),$$

which is an equilibrium point of system (28)-(29) and corresponds to the equilibrium point $p_2$ of system (7)-(9) restricted to $\mathcal{A}$.

Let us distinguish the following cases:

(1) Case $\beta < (\delta+\mu)(\mu+p)\mu^{-1}$. In this case, there are three regions (see Figure 4):

$$\mathcal{R}_1 = \left\{(S,I)\in\mathbb{R}^2 \,:\, S>0,\ I>0,\ S>\frac{\nu(\delta+\mu)}{\beta\mu}\right\},$$
$$\mathcal{R}_2 = \left\{(S,I)\in\mathbb{R}^2 \,:\, S>0,\ I>0,\ S<\frac{\nu(\delta+\mu)}{\beta\mu},\ I>\frac{\nu}{\beta\mu}\left(\frac{\nu}{S}-(\mu+p)\right)\right\},$$
$$\mathcal{R}_3 = \left\{(S,I)\in\mathbb{R}^2 \,:\, S>0,\ I>0,\ I<\frac{\nu}{\beta\mu}\left(\frac{\nu}{S}-(\mu+p)\right)\right\}.$$

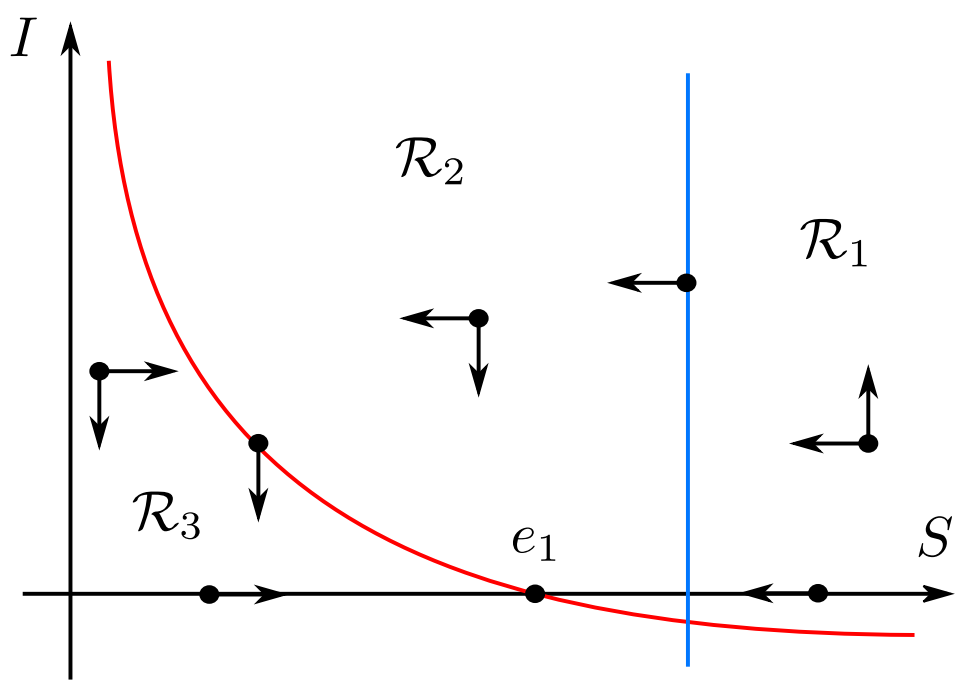

Figure 4: Nullclines, regions and vector field in case $\beta < (\delta+\mu)(\mu+p)\mu^{-1}$.

Assume that a solution starts in $\mathcal{R}_1$. Then, from (28)-(29), $S' < 0$ and $I' > 0$, whence $S$ decreases and $I$ increases. Next, the solution intersects the $I-$nullcline (31) (the solution is bounded since $\mathcal{A}$ is a compact set) and enters $\mathcal{R}_2$. By a similar reasoning, we obtain the arrows in Figure 4 indicating the vector field in every region or nullcline, whence we deduce that any solution starting in $\mathcal{A}$ converges to $e_1$.

(2) Case $\beta = (\delta+\mu)(\mu+p)\mu^{-1}$. In this case, we only have two regions. By a similar reasoning to the previous case, we obtain that every solution converges to $e_1$.

(3) Case $\beta > (\delta+\mu)(\mu+p)\mu^{-1}$. In this case we have the following four regions (see Figure 5):

$$\mathcal{R}_1 = \left\{(S,I)\in\mathbb{R}^2 \,:\, S>0,\ I>0,\ S>\frac{\nu(\delta+\mu)}{\beta\mu},\ I>\frac{\nu}{\beta\mu}\left(\frac{\nu}{S}-(\mu+p)\right)\right\},$$
$$\mathcal{R}_2 = \left\{(S,I)\in\mathbb{R}^2 \,:\, S>0,\ I>0,\ S<\frac{\nu(\delta+\mu)}{\beta\mu},\ I>\frac{\nu}{\beta\mu}\left(\frac{\nu}{S}-(\mu+p)\right)\right\},$$
$$\mathcal{R}_3 = \left\{(S,I)\in\mathbb{R}^2 \,:\, S>0,\ I>0,\ S<\frac{\nu(\delta+\mu)}{\beta\mu},\ I<\frac{\nu}{\beta\mu}\left(\frac{\nu}{S}-(\mu+p)\right)\right\},$$
$$\mathcal{R}_4 = \left\{(S,I)\in\mathbb{R}^2 \,:\, S>0,\ I>0,\ S>\frac{\nu(\delta+\mu)}{\beta\mu},\ I<\frac{\nu}{\beta\mu}\left(\frac{\nu}{S}-(\mu+p)\right)\right\}.$$

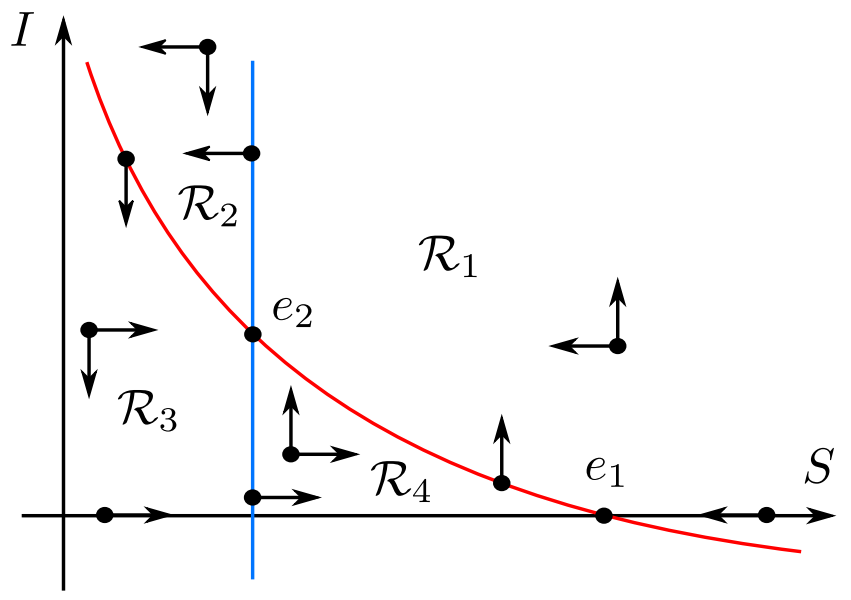

Figure 5: Nullclines, regions and vector field in case $\beta > (\delta + \mu)(\mu + p)\mu^{-1}$.

Similarly to the study in the first case, we obtain the arrows in Figure 5 indicating the vector field. Nevertheless, the dynamics now is more complicated.

Let us suppose that a solution starts in $\mathcal{R}_1$. Then, either it intersects the $I-$nullcline (31) and enters $\mathcal{R}_2$ or it converges to $e_2$. If the solution enters $\mathcal{R}_2$, it intersects the $S-$nullcline (32) and enters $\mathcal{R}_3$. When a solution enters $\mathcal{R}_3$, either it converges to $e_2$ or it intersects the $I-$nullcline (31) and enters $\mathcal{R}_4$, from which the solution intersects the $S-$nullcline (32) and enters $\mathcal{R}_1$.

Notice that nothing prevents our solution from oscillating indefinitely around $e_2$, for instance, by converging to a possible periodic orbit of the system in $\mathcal{A}$. Hence, consider the Dulac function $D(S, I) = 1/I$, where $D \in \mathcal{C}^1(\mathbb{R}^2_+)$. Then, it yields that

$$\frac{\partial(fD)}{\partial S} + \frac{\partial(gD)}{\partial I} = -\left[\frac{\beta\mu}{\nu} + \frac{\mu + p}{I}\right] < 0,$$

where $f$ and $g$ denote the right-hand side of (28) and (29), respectively. Hence, by combining the Poincaré-Bendixson Trichotomy and the Dulac-Bendixson Criterion (see [4, 17]), there are no periodic orbits for (28), i.e., every solution in $\mathcal{A}$ converges to $e_2$ (unless $I_0 = 0$ in which case it converges to $e_1$).

□

**Remark 4.3.** *From Theorem 4.2, we can deduce that the disease becomes endemic as long as condition* (20) *holds true, and the disease is eradicated otherwise.*

### 4.2. Additional global results concerning endemic diseases

In this section we stablish some additional results concerning the case in which the disease becomes endemic, provided that inequality (20) is fulfilled, regardless the initial conditions are not in the attracting set $\mathcal{A}$. To this end, first we need to prove two preliminary lemmas.

**Lemma 4.1.** *Let* $(S(t;0,S_0), I(t;0,I_0), R(t;0,R_0))$ *be a solution to* (7)-(9) *with initial condition* $(S_0, I_0, R_0) \in \mathbb{R}^3_+$. *Then, it satisfies*

$$S'(t) = \nu - \frac{\beta}{De^{-\mu t} + \frac{\nu}{\mu}}\, S(t)I(t) - (\mu + p)\, S(t), \tag{33}$$

$$I'(t) = \left(\frac{\beta}{De^{-\mu t} + \frac{\nu}{\mu}}\, S(t) - (\delta + \mu)\right) I(t), \tag{34}$$

*where* $D = (S_0 + I_0 + R_0) - \nu/\mu = N_0 - \nu/\mu$.

*Proof.* Defining again the total population $N(t) = S(t) + I(t) + R(t)$, it satisfies $N'(t) = -\mu N(t) + \nu$, whose solution is given by

$$N(t;0,N_0) = De^{-\mu t} + \frac{\nu}{\mu}, \tag{35}$$

with $N_0 = D + \nu/\mu$. Therefore, it remains to substitute (35) in (7)-(9) to conclude. □

**Lemma 4.2.** *Let* $(S(t;0,S_0), I(t;0,I_0), R(t;0,R_0))$ *be a solution to* (7)-(9) *with initial condition* $S(0) > 0$, $I(0) > 0$ *and* $R(0) > 0$. *Assume that*

$$C = \left\{ t \,:\, S(t) > \frac{(\delta+\mu)\left(De^{-\mu t} + \frac{\nu}{\mu}\right)}{\beta} \right\} \neq \emptyset.$$

*Then, for every interval* $J \subset C$ *it is satisfied that* $I$ *is strictly increasing in* $J$.

*Proof.* First note that, by the continuity of $S$ and condition $C \neq \emptyset$, there exist infinite intervals $J$ such that $J \subset C$.

By Lemma 4.1, we have

$$I'(t) = \frac{\beta}{De^{-\mu t} + \frac{\nu}{\mu}} S(t)I(t) - (\delta+\mu)I(t) = \left(\frac{\beta}{De^{-\mu t} + \frac{\nu}{\mu}} S(t) - (\delta+\mu)\right) I(t).$$

Since $I_0 > 0$, we have that $I(t;0,I_0) > 0$ for every $t$ by Lemma 3.1. On the other hand, for every interval $J$ such that $J \subset C$ we have that

$$S(t;0,S_0) > \frac{(\delta+\mu)(De^{-\mu t} + \frac{\nu}{\mu})}{\beta}$$

for every $t \in J$ and $S_0 > 0$, which implies that

$$\frac{\beta}{De^{-\mu t} + \frac{\nu}{\mu}} S(t) - (\delta+\mu) > 0$$

for every $t \in J$, then

$$I'(t) = (\frac{\beta}{De^{-\mu t} + \frac{\nu}{\mu}} S - (\delta + \mu))I(t) > 0$$

for every $t \in J$ and, then, $I$ is strictly increasing in $J$ as desired. $\square$

Now we state the main result of this section.

**Proposition 4.4.** *Let* $(S(t; 0, S_0), I(t; 0, I_0), R(t; 0, R_0))$ *be a solution to* (7)-(9) *with initial condition* $S(0) > 0$, $I(0) > 0$ *and* $R(0) > 0$ *and assume that* (20) *fulfills. Then, it is satisfied that* $I(t) \nrightarrow 0$ *as* $t \to \infty$.

*Proof.* Assume that $I(t) \to 0$ as $t \to +\infty$. Then, adding (7) and (8), we have that

$$S'(t) + I'(t) = -(\mu + p)S(t) + \nu - (\delta + \mu)I(t) = -(\mu + p)(S(t) + I(t)) + \nu + (p - \delta)I(t).$$

By Lemma 5.1 (we prefer to keep the mentioned lemma after this proposition not to make the organization of the paper more complicated), $S(t) + I(t) \to \nu/(\mu + p)$ as $t \to +\infty$, whence $S(t) \to \nu/(\mu + p)$ as $t \to +\infty$.

Therefore, if there exists $t_0 \geq 0$ such that $S(t; 0, S_0) > (\delta + \mu)(De^{-\mu t} + \frac{\nu}{\mu})\beta^{-1}$ for every $t > t_0$, we have that $I(t; 0, I_0)$ is a positive and strictly increasing function for $t > t_0$ by Lemma 4.2, which contradicts that $I(t) \to 0$. Thus, there exists a sequence $t_n \to +\infty$ as $n \to +\infty$ such that

$$S(t_n) \leq \frac{(\delta + \mu)(De^{-\mu t_n} + \frac{\nu}{\mu})}{\beta}$$

for every $n$ and, by taking limits, we obtain

$$\lim_{n \to +\infty} S(t_n) = \frac{\nu}{\mu + p} \leq \lim_{n \to +\infty} \frac{(\delta + \mu)(De^{-\mu t_n} + \frac{\nu}{\mu})}{\beta} = \frac{\nu(\delta + \mu)}{\mu\beta},$$

then $\beta \leq (\delta + \mu)(p + \mu)\mu^{-1}$, a contradiction with (20), whence $I(t) \nrightarrow 0$ as $t \to +\infty$. $\square$

**Remark 4.4.** *We would like to point out that Proposition 4.4 carries out that the infection does not become extinct if inequality* (20) *is satisfied, regardless the initial conditions.*

### 4.3. Numerical simulations

In this section, we depict some numerical simulations aimed at illustrating the theoretical results established throughout this section. Each figure consists of a panel on the right, which displays the phase portrait of the system with an arrow indicating the initial condition, and three smaller panels on the left, where the time evolution of the susceptible, infected, and recovered individuals is plotted.

In Figure 6 we consider $\nu = 5$, $\beta = 1.5$, $\mu = 0.5$, $\delta = 0.7$, $p = 0.4$ and the initial condition $(S_0, I_0, R_0) = (25, 2, 0)$. Since (24) fulfills, the disease is eradicated, as we already proved.

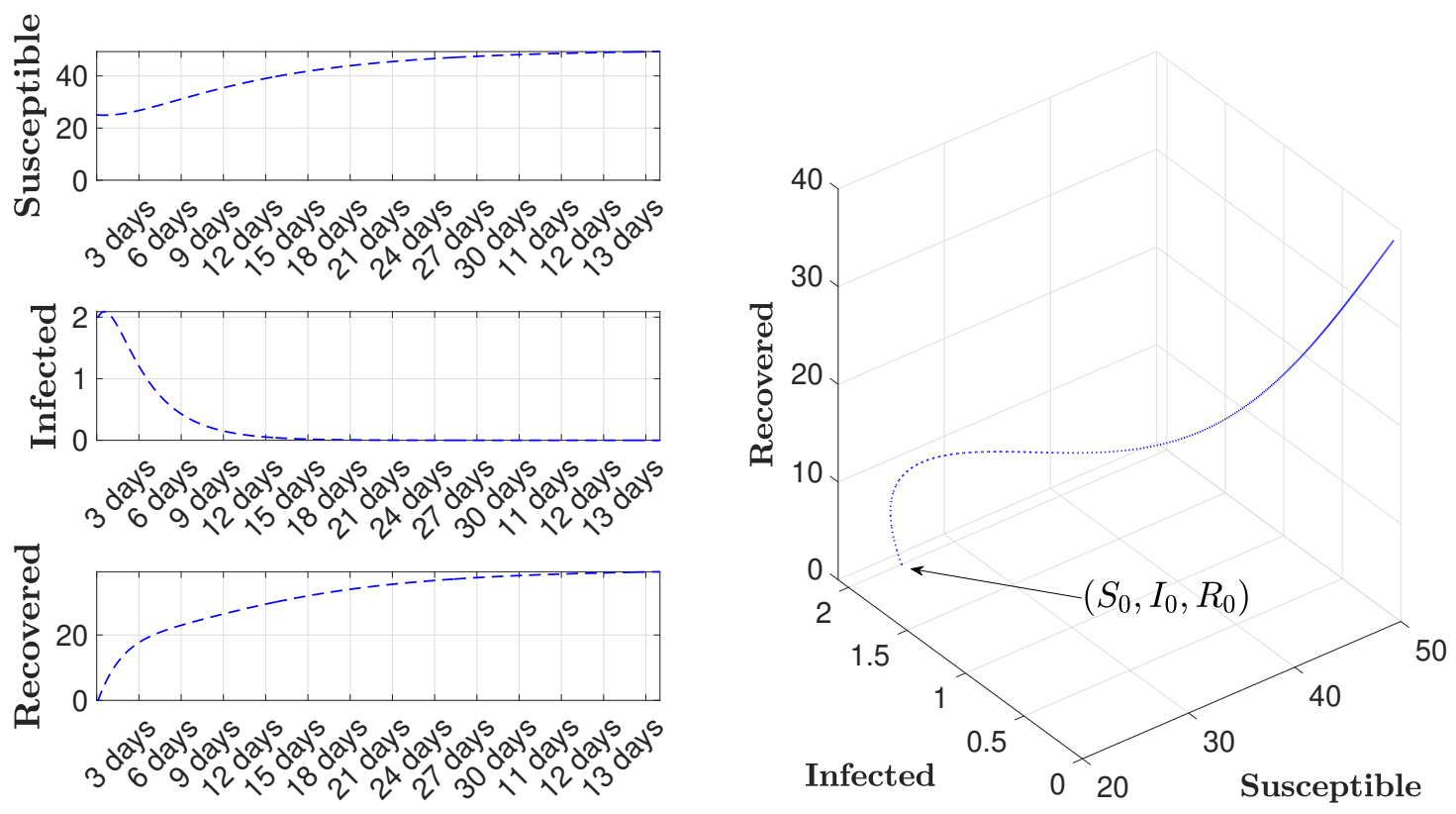

Figure 6: **The disease is eradicated** with $\nu = 5$, $\beta = 1.5$, $\mu = 0.5$, $\delta = 0.7$, $p = 0.4$ and the initial condition $(S_0, I_0, R_0) = (25, 2, 0)$.

Nevertheless, in Figure 7 we increase the value of $\beta = 2.4$. As a consequence, (20) fulfills and the disease becomes endemic, as we proved before.

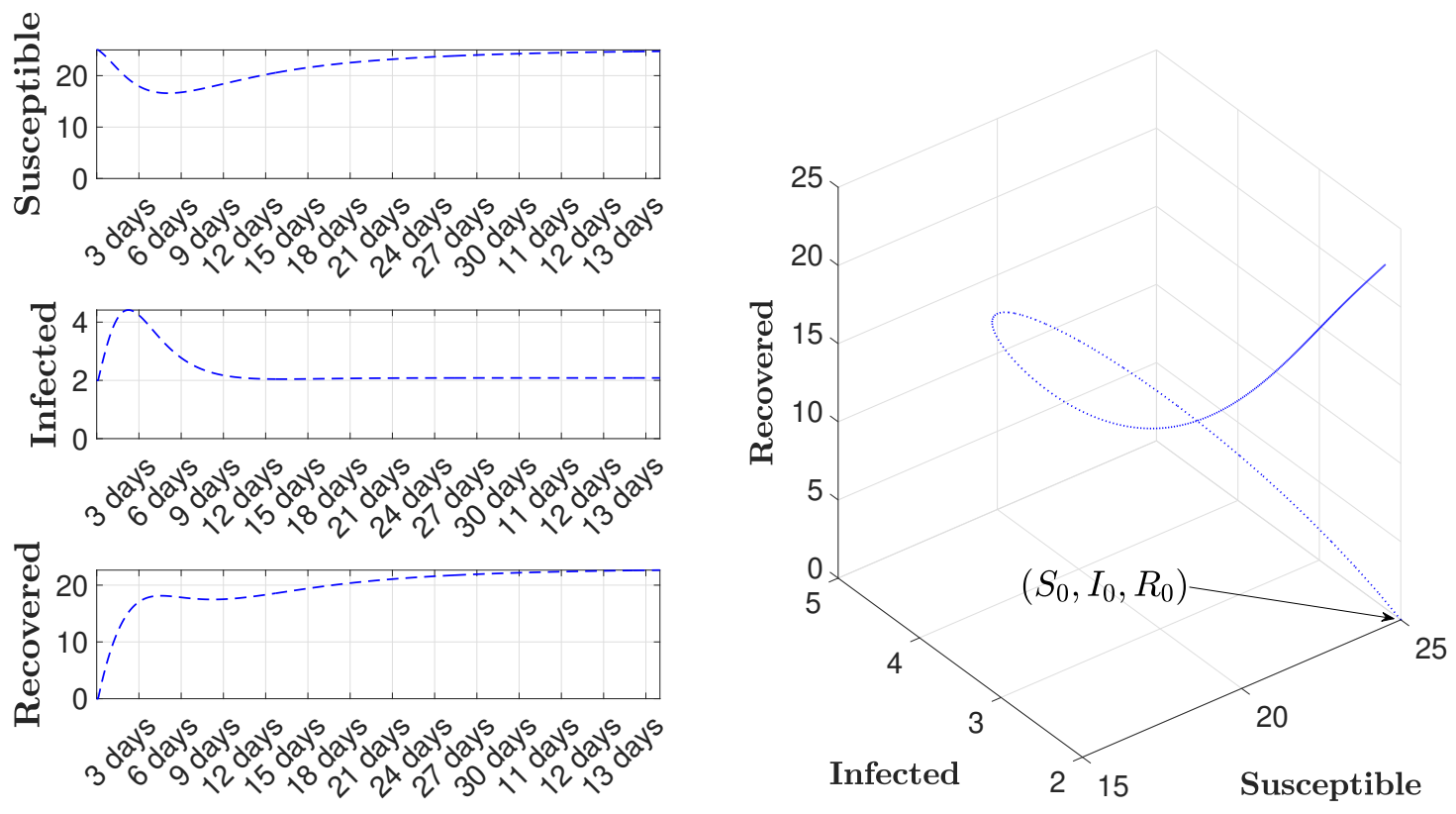

Figure 7: **The disease becomes endemic** with $\nu = 5$, $\beta = 2.4$, $\mu = 0.5$, $\delta = 0.7$, $p = 0.4$ and the initial condition $(S_0, I_0, R_0) = (25, 2, 0)$.

## 5. SIR model with demography, random transmission coefficient and non-autonomous vaccination rate

In this section we investigate the SIR model with demography, random transmission coefficient and non-autonomous vaccination rate (10)-(12). The existence and uniqueness of non-negative global solution of the system was proved in Theorem 3.1. Our aim now is to analyze in detail the long-time behavior of the solutions of the system, i.e., to provide conditions on the parameters under which the disease is eradicated or becomes endemic. Nevertheless, we need to include a brief subsection with a technical result first.

### 5.1. A previous technical result

In this section, we prove a technical result concerning supersolutions of differential systems, which will be useful later to obtain upper bounds of the solutions of the systems under study.

Let us start by defining the concept of supersolution.

**Definition 5.1.** *It is said that* $\big(\overline{S}(t), \overline{I}(t), \overline{R}(t)\big)$ *is a supersolution of system* (10)-(12) *if we have that* $S(t) \leq \overline{S}(t)$, $I(t) \leq \overline{I}(t)$ *and* $R(t) \leq \overline{R}(t)$ *are fulfilled for all* $t \geq 0$*.*

Now, we present the technical result we need in the sequel.

**Lemma 5.1.** *Consider the following Cauchy problem*

$$x'(t) = a(t) - b(t)x(t), \tag{36}$$

$$x(\tau) = x_\tau, \tag{37}$$

*where* $b(t) \geq \alpha > 0$*, and define the function* $\ell(t) = a(t)/b(t)$*. Hence,* $x(t) \to \sigma$ *when* $t \to +\infty$ *provided that* $l(t) \to \sigma$ *is fulfilled when* $t \to +\infty$*.*

*Proof.* Following the ideas in [14], we can rewrite the Cauchy problem (36)-(37) as

$$x'(t) = b(t)(\ell(t) - x(t)), \tag{38}$$

$$x(\tau) = x_\tau, \tag{39}$$

whose solution is given by

$$x(t; \tau, x_s) = x_s e^{-\int_s^t b(\theta)d\theta} + \int_s^t \ell(\theta)b(\theta)e^{-\int_\theta^t b(\tau)d\tau}d\theta,$$

for every $t \geq \tau$.

Since $\ell(t) \to \sigma$ when $t \to +\infty$, for any given $\varepsilon > 0$, there exists $t_\varepsilon > \tau$ such that $\ell(t) > \sigma - \varepsilon$ for all $t > t_\varepsilon$.

Let us now consider the following Cauchy problem

$$x'(t) = b(t)(\sigma_{\varepsilon^-} - x(t)), \tag{40}$$
$$x(t_\varepsilon) = x_\varepsilon, \tag{41}$$

where we denote $\sigma_{\varepsilon^-} = \sigma - \varepsilon$ and $x_\varepsilon = x(t_\varepsilon; \tau, x_\tau)$.

Without loss of generality, let us assume $x(t_\varepsilon; \tau, x_\tau) < \sigma$. In fact, the solution $x_{\sigma_{\varepsilon^-}}(t; t_\varepsilon, x_\varepsilon)$ of (40) is always below $x(t; t_\varepsilon, x_\varepsilon)$, since $\sigma_{\varepsilon^-} \leq \ell(t)$ for all $t > t_\varepsilon$. Indeed,

$$x_{\sigma_{\varepsilon^-}}(t; t_\varepsilon, x_\varepsilon) = x_\varepsilon e^{-\int_{t_\varepsilon}^t b(\theta)d\theta} + \int_{t_\varepsilon}^t \sigma_{\varepsilon^-} b(\theta) e^{-\int_\theta^t b(\tau)d\tau} d\theta,$$

whence we obtain that

$$x(t; t_\varepsilon; x_\varepsilon) - x_{\sigma_{\varepsilon^-}}(t; t_\varepsilon, x_\varepsilon) = \int_{t_\varepsilon}^t (\ell(\theta) - \sigma_{\varepsilon^-}) b(\theta) e^{-\int_\theta^t b(\tau)d\tau} d\theta \geq 0,$$

for all $t > t_\varepsilon$, thanks to the fact that $b(t) > 0$ in $\mathbb{R}$.

On another hand, it yields that

$$x_{\sigma_\varepsilon^-}(t; t_\varepsilon, x_\varepsilon) = x_\varepsilon e^{-\int_{t_\varepsilon}^t b(\theta)d\theta} + \sigma_\varepsilon^- \left(1 - e^{-\int_{t_\varepsilon}^t b(\theta)d\theta}\right) \to \sigma_{\varepsilon^-}$$

when $t \to +\infty$.

Analogously, it can be proved that the solution of the Cauchy problem

$$x'(t) = b(t)(\sigma_{\varepsilon^+} - x(t)), \tag{42}$$
$$x(\tau_\varepsilon) = x_\varepsilon, \tag{43}$$

where $\sigma_{\varepsilon^+} = \sigma + \varepsilon$, $x_\varepsilon = x(\tau_\varepsilon; s, x_s)$, and $\tau_\varepsilon$ is such that $\ell(t) \leq \sigma + \varepsilon$ for all $t > \tau_\varepsilon$, is always over $x(t; \tau_\varepsilon, x_\varepsilon)$ and converges to $\sigma_{\varepsilon^+}$ when $t \to +\infty$.

Since $\varepsilon > 0$ can be arbitrarily small, we conclude that any solution of (36) converges to $\sigma$ when $t \to +\infty$. □

### 5.2. Long-time dynamics of the solutions

This section is dedicated to study the long-time dynamics of the solutions of system (10)-(12), i.e., we aim to provide conditions on the parameters of the systems under which the disease either is eradicated or becomes endemic.

Consider the following Cauchy problem

$$\overline{S}'(t) = \nu - (\mu + \overline{p})\, \overline{S}(t), \tag{44}$$

$$\overline{I}'(t) = \left( (\beta + d)\, \frac{\overline{S}(t,\tau)}{N(t,\tau)} - (\delta + \mu) \right) \overline{I}(t), \tag{45}$$

$$\overline{R}'(t) = \delta\, \overline{I}(t) + \overline{p}\, \overline{S}(t) - \mu\, \overline{R}(t), \tag{46}$$

$$\overline{S}(0) = S_0,\ \overline{I}(0) = I_0,\ \overline{R}(0) = R_0, \tag{47}$$

where $(S_0, I_0, R_0) \in \mathbb{R}^3_+$.

As in Section 4, we can consider here the total population $N$ and it yields that

$$\lim_{t-\tau\to+\infty} N(t : 0, \omega, N_0) = \frac{\nu}{\mu},$$

for any $\omega \in \Omega$ and $N_0 = S_0 + I_0 + R_0 \geq 0$. Thus, the compact set $\mathcal{A}$ given by (27) is also an invariant and attracting set for the solutions of system (10)-(12). Notice that, since $\mathcal{A}$ is deterministic, it is worthy to point out that does not depend on the noise.

Moreover, we have that $S(t; 0, \omega, S_0) \leq \overline{S}(t; 0, \omega, S_0)$, $I(t; 0, \omega, I_0) \leq \overline{I}(t; 0, \omega, I_0)$ and $R(t; 0, \omega, R_0) \leq \overline{R}(t; 0, \omega, R_0)$.

**Theorem 5.1.** *Consider system* (10)-(12) *with initial condition* $(S_0, I_0, R_0) \in \mathbb{R}^3_+$ *and define* $\overline{\rho}_0 = (\delta + \mu)(\mu + \overline{p})\mu^{-1}$. *Then, as long as*

$$\beta < \overline{\rho}_0 - d \tag{48}$$

*holds true, we have that* $I(t) \to 0$ *when* $t \to +\infty$. *In addition, for any* $\omega \in \Omega$, *every solution of Cauchy problem* (44)-(47) *verifies that*

$$\overline{S}(t; 0, \omega, S_0) \to \frac{\nu}{\mu + \overline{p}}, \quad \overline{I}(t; 0, \omega, I_0) \to 0, \quad \text{and} \quad \overline{R}(t; 0, \omega, R_0) \to \frac{\nu\, \overline{p}}{\mu(\mu + \overline{p})}$$

*when* $t \to +\infty$.

*Proof.* It is straightforward to prove that, for any $\omega \in \Omega$ and $S_0 \geq 0$,

$$\overline{S}(t; 0, \omega, S_0) \to \frac{\nu}{\mu + \overline{p}}$$

when $t \to +\infty$. Then, since for any $\omega \in \Omega$ and $N_0 \geq 0$ we have

$$N(t; 0, \omega; N_0) \to \frac{\nu}{\mu},$$

when $t \to +\infty$, then, for any $\omega \in \Omega$ and $(S_0, I_0, R_0) \in \mathbb{R}^3_+$, it yields that

$$\frac{\overline{S}(t; 0, \omega, S_0)}{N(t; 0, \omega, N_0)} \to \frac{\mu}{\mu + \overline{p}} = \frac{\delta + \mu}{\overline{\rho}_0} \tag{49}$$

when $t \to +\infty$.

Now, consider $\varepsilon_0 = (\beta + d)/\overline{\rho}_0 < 1$, whence $\beta = \varepsilon_0\ \overline{\rho}_0 - d$. Therefore, for any $\omega \in \Omega$, $S_0 \geq 0$ and $N_0 \geq 0$, we obtain that

$$(\beta + d)\ \frac{\overline{S}(t; 0, \omega, S_0)}{N(t; 0, \omega, N_0)} - (\delta + \mu) = \varepsilon_0\ \overline{\rho}_0\ \frac{\overline{S}(t; 0, \omega, S_0)}{N(t; 0, \omega, N_0)} - (\delta + \mu).$$

On the other hand, thanks to (49), for any given $\varepsilon > 0$, $\omega \in \Omega$, $S_0 \geq 0$ and $N_0 \geq 0$, there exists $t_\varepsilon > 0$ such that

$$\frac{\overline{S}(t; 0, \omega, S_0)}{N(t; 0, \omega, N_0)} < \frac{\delta + \mu}{\overline{\rho}_0} + \varepsilon$$

for all $t > t_\varepsilon$, from which we obtain that

$$\begin{aligned}\varepsilon_0\ \overline{\rho}_0\ \frac{\overline{S}(t; 0, \omega, S_0)}{N(t; 0, \omega, N_0)} - (\delta + \mu) &< \varepsilon_0\ \overline{\rho}_0\ \left(\frac{\delta + \mu}{\overline{\rho}_0} + \varepsilon\right) - (\delta + \mu) \\ &= \varepsilon_0\ (\delta + \mu) + \varepsilon_0\ \varepsilon\ \overline{\rho}_0 - (\delta + \mu) \\ &= (\varepsilon_0 - 1)\ (\delta + \mu) + \varepsilon_0\ \varepsilon\ \overline{\rho}_0,\end{aligned}$$

for any given $\varepsilon > 0$, $\omega \in \Omega$, $S_0 \geq 0$, $N_0 \geq 0$ and for all $t > t_\varepsilon$.

Denoting now $a = (\varepsilon_0 - 1) + \varepsilon_0\ \varepsilon\ \overline{\rho}_0$ and considering $\varepsilon < \dfrac{(1 - \varepsilon_0)(\delta + \mu)}{\varepsilon_0\ \overline{\rho}_0}$, we have that $\overline{I}'(t) < a\overline{I}(t)$ for $t > t_\varepsilon$, where $a < 0$. Then,

$$I(t; t_\varepsilon, \omega, I(t_\varepsilon)) \leq \overline{I}(t; t_\varepsilon, \omega, I(t_\varepsilon)) \leq I(t_\varepsilon; t_\varepsilon, \omega, I(t_\varepsilon)) e^{a(t - t_\varepsilon)} \to 0$$

when $t \to +\infty$.

Finally, in order to study the limit of $\overline{R}(t; 0, \omega, R_0)$, we can define the function

$$\ell_{\overline{R}}(t) = \frac{\delta\ \overline{I}(t; 0, \omega, I_0) + \overline{p}\ \overline{S}(t; 0, \omega, S_0)}{\mu}.$$

Due to the convergence of $\overline{I}(t; 0, \omega, I_0)$ and $\overline{S}(t; 0, \omega, S_0)$, we have that

$$\ell_{\overline{R}}(t) \to \frac{\nu\ \overline{p}}{\mu\ (\mu + \overline{p})}$$

when $t \to +\infty$. Thus, by Lemma 5.1, we can conclude that

$$\overline{R}(t; 0, \omega, R_0) \to \frac{\nu\ \overline{p}}{\mu\ (\mu + \overline{p})},$$

when $t \to +\infty$, which completes the proof. □

Next, we provide conditions on the transmission coefficient of the disease under which the disease becomes endemic.

**Theorem 5.2.** *Assume that*

$$\beta \geq \frac{\nu\,(\delta+\mu)\,(\mu+m)}{\mu\,\nu} + d \tag{50}$$

*holds true, where* $m = \max\{\delta, \overline{p}\}$. *Then, the number of infected individuals does not converge to zero.*

*Proof.* From (10) and (11), it yields that

$$(S(t)+I(t))' \geq \nu - (\mu+m)(S(t)+I(t)),$$

whence, for a given $\varepsilon_0 > 0$, $\omega \in \Omega$, $S_0 > 0$ and $I_0 > 0$, there exists $t_{\varepsilon_0} > 0$ such that

$$S(t;0,\omega,S_0) + I(t;0,\omega,I_0) > \frac{\nu}{\mu+m} - \varepsilon_0 \tag{51}$$

for all $t > t_{\varepsilon_0}$.

On another hand, since $N(t;0,\omega,N_0) \to \dfrac{\nu}{\mu}$ for any $\omega \in \Omega$ and $N_0 > 0$ when $t \to +\infty$, for any given $\varepsilon_1 > 0$, $\omega \in \Omega$ and $N_0 > 0$, there exists $t_{\varepsilon_1} > 0$ such that

$$\frac{1}{N(t;0,\omega,N_0)} > \frac{\mu}{\nu} - \varepsilon_1, \tag{52}$$

for $t > t_{\varepsilon_1}$.

Consider now $t_\varepsilon > \max\{t_{\varepsilon_0}, t_{\varepsilon_1}\}$ and let $\varepsilon > 0$ be such that $I(t_\varepsilon;0,\omega,I_0) = \varepsilon$. Hence, evaluating (11) at $t = t_\varepsilon$, we have that

$$\left.\frac{dI(t)}{dt}\right|_{t=t_\varepsilon} \geq \left((\beta-d)\frac{S(t_\varepsilon)}{N(t_\varepsilon)} - (\delta+\mu)\right)\varepsilon.$$

Now, since $t_\varepsilon > \max\{t_{\varepsilon_0}, t_{\varepsilon_1}\}$, thanks to (51) and (52), we obtain that

$$(\beta-d)\frac{S(t_\varepsilon;0,\omega,S_0)}{N(t_\varepsilon;0,\omega,N_0)} - (\delta+\mu) > (\beta-d)\left(\frac{\nu}{\mu+m} - \varepsilon_0 - \varepsilon\right)\left(\frac{\mu}{\nu} - \varepsilon_1\right) - (\delta+\mu),$$

for any $\omega \in \Omega$, $S_0 > 0$ and $N_0 > 0$.

Thus, from (50) we have that $I'(t_\varepsilon;0,\omega,I_0) > 0$ for any $\omega \in \Omega$ and $I_0 > 0$, i.e., $I$ increases at $t_\varepsilon$. Moreover, this happens as soon as $I$ reaches the value $\varepsilon$. Therefore, $I$ does not converge to zero and remains always strictly positive. □

**Remark 5.1.** *It is worth noting that Theorem 5.1 and Theorem 5.2 establish conditions under which the disease is eradicated or becomes endemic. In essence, the results indicate that the transmission coefficient must remain below a certain threshold (as specified in* (48)*) to guarantee eradication. On the other hand, if the transmission coefficient is large enough (more precisely, if* (50) *holds true), the disease becomes endemic.*

### 5.3. Numerical simulations

In this section we display some numerical simulations aimed at illustrating the theoretical results established throughout the paper.

Each figure consists of a panel on the right, which shows the phase portrait of the system with an arrow indicating the initial condition, and three smaller panels on the left, where the time evolution of the susceptible, infected, and recovered populations is plotted. The blue dashed/dotted line corresponds to the solution of the system in the absence of stochastic perturbations, while the solid colored lines represent different realizations of the random models.

In Figure 8 we consider $\nu = 5$, $\beta = 1$, $\mu = 0.5$, $\delta = 0.7$, $p = 0.4$, $d = 1.5$, $p(t) = 0.1(1 + \sin(t/(1 + |t|)))$ and the initial condition $(S_0, I_0, R_0) = (25, 2, 0)$. Since (48) is fulfilled, the disease is eradicated, as we already proved.

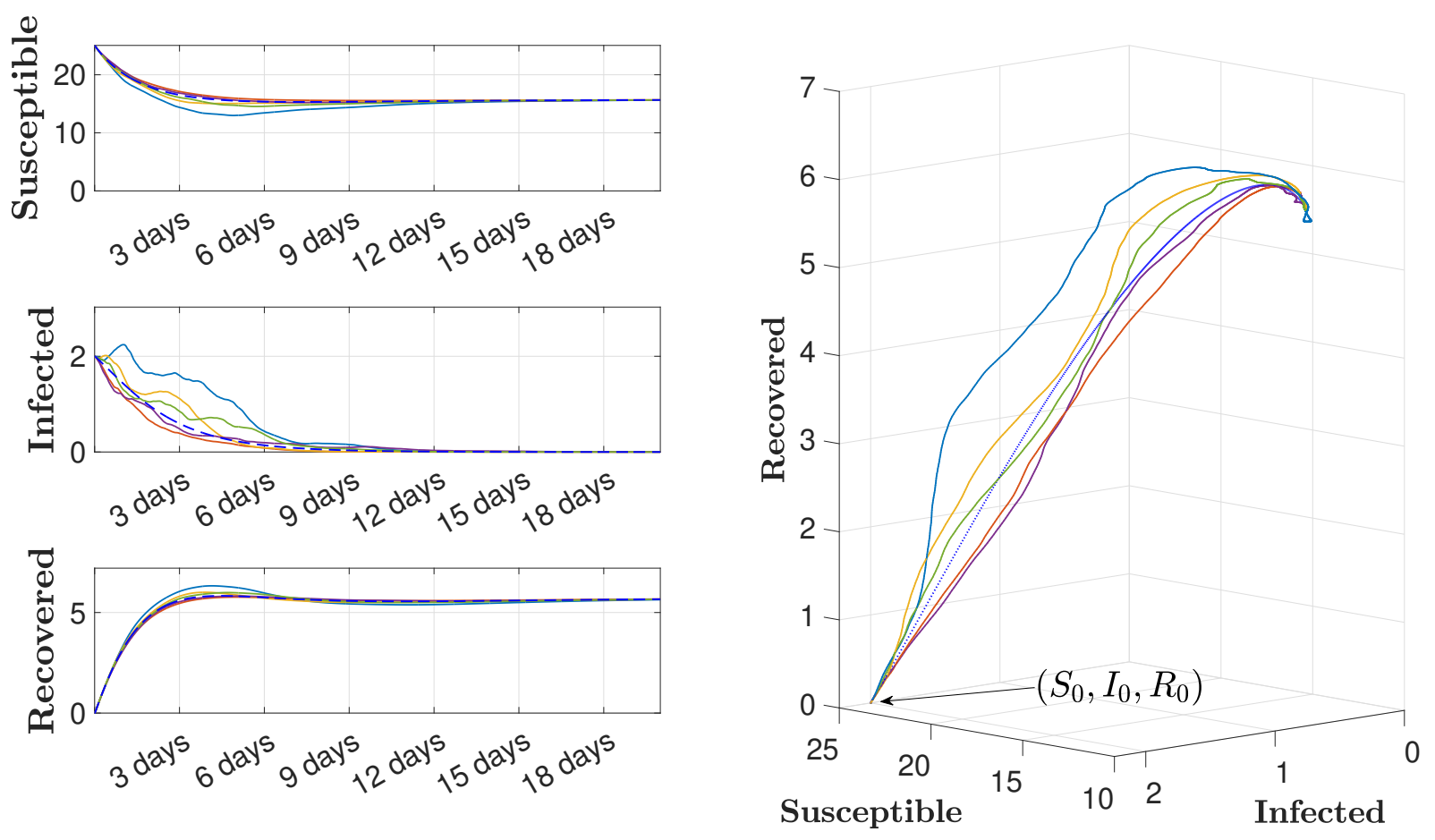

Figure 8: **The disease is eradicated** with $\nu = 5$, $\beta = 1$, $\mu = 0.5$, $\delta = 0.7$, $p(t) = 0.1(1 + \sin(t/(1 + |t|)))$, $d = 1.5$ and the initial condition $(S_0, I_0, R_0) = (25, 2, 0)$.

Nevertheless, in Figure 9 we increase the value of $\beta = 5.5$. As a consequence, (50) fulfills and the disease becomes endemic, as we proved before.

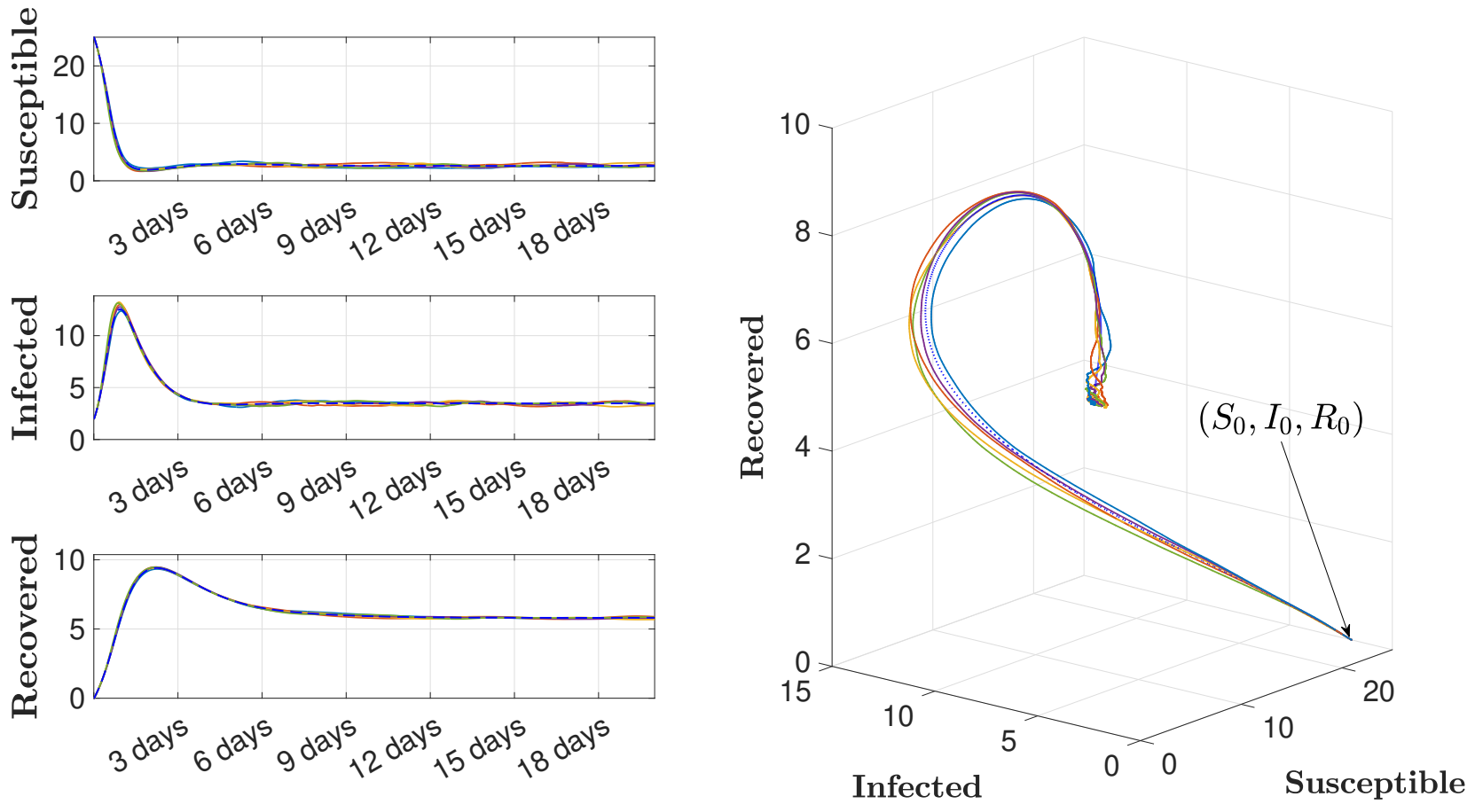

Figure 9: **The disease becomes endemic** with $\nu = 5$, $\beta = 5.5$, $\mu = 0.5$, $\delta = 0.7$, $p(t) = 0.1(1 + \sin(t/(1 + |t|)))$, $d = 1.5$ and the initial condition $(S_0, I_0, R_0) = (25, 2, 0)$.

## 6. Conclusions and final comments

In this work we address an extension of the classical SIR model incorporating demography, a time-dependent vaccination rate of susceptible individuals and bounded random perturbations on the transmission coefficient, aiming to develop more realistic frameworks for the mathematical modeling of epidemics.

We start by analyzing in detail the autonomous version of the model. Specifically, we compute its equilibrium points, investigate their local stability, and prove the existence of an invariant and attracting set for the system. After that, we focus on characterizing the global asymptotic behavior of the solutions restricted to this attracting set. Moreover, we derive the corresponding basic reproduction number.

Motivated by real-world scenarios, we examine in depth the random system. Particular attention is given to the impact of the random perturbations on the transmission coefficient, as these are capable of modeling natural movement patterns of individuals, which plays a fundamental role in the dynamics of the solutions, as our analysis reveals.

In both models we are able to provide conditions on the parameters of the systems under which the disease either is eradicated or becomes endemic, which is the most important goal in real applications.

In summary, this paper sheds light on new aspects of epidemic modeling, including vaccination strategies and random terms that can affect not only to the transmission coefficient, but also other parameters and alternative formulations of classical models.

**Financial disclosure**

This work has been partially supported by the Spanish Ministerio de Ciencia e Innnovación, Agencia Estatal de Investigación (AEI) and Fondo Europeo de Desarrollo Regional (FEDER) under the project PID2021-122991NB-C21, by the Spanish Ministerio de Economía y Competitividad of Spain under grant PID2019-104735RB-C42 and by the Madrid Government (Comunidad de Madrid-Spain) under the Multiannual Agreement 2023-2026 with Universidad Politécnica de Madrid in the Line A, Emerging PhD researchers, grant agreement DOCTORES-EMERGENTES-24-UMDLRU-108-XVLWPC.

**Conflict of interest**

The authors declare no potential conflict of interests.